\documentclass[preprints,article,accept,oneauthor,pdftex]{Definitions/mdpi} 

\firstpage{1}
\usepackage{graphicx}
\usepackage{bm,amsfonts,amsthm,placeins}
\newcommand{\gl}[1]{(\ref{#1})}

\pubvolume{xx}
\articlenumber{5}
\pubyear{2019}
\copyrightyear{2019}
\history{Received: date; Accepted: date; Published: date}

\Title{Magnetic Structure of CoO} 
  
\Author{Ekkehard Kr\"uger \orcidA{}}

\AuthorNames{Firstname Lastname, Firstname Lastname and Firstname Lastname}

\address[1]{%
Institut f{\"u}r Materialwissenschaft, Materialphysik,
  Universit{\"a}t Stuttgart, D-70569 Stuttgart, Germany}

\corres{Correspondence: ekkehard.krueger@imw.uni-stuttgart.de}

\abstract{The paper reports theoretical evidence that the
  multi-spin-axis magnetic structure proposed in 1964 by van Laar
  actually does exist in CoO. This tetragonal spin arrangement
  produces both the strong tetragonal and the weaker monoclinic
  distortion experimentally observed in this material.  The monoclinic
  distortion is proposed to be a ``monoclinic-like'' distortion of the
  array of the oxygen atoms, comparable with the rhombohedral-like
  distortion of the oxygen atoms recently proposed to be present in
  NiO and MnO. The monoclinic-like distortion has no influence on the
  tetragonal magnetic structure which is generated by a special
  nonadiabatic atomic-like motion of the electrons near the Fermi
  level. It is argued that it is this special atomic-like motion which
  qualifies CoO to be a Mott insulator.}

\begin{document}


\section{Introduction}
Cobalt monoxide is antiferromagnetic with the N\'eel temperature
$T_N = 289$ K. Just as the other isomorphic transition-metal monoxides
MnO, FeO, and NiO, it is a Mott insulator in both, the paramagnetic
and the antiferromagnetic phase. While above $T_N$, all the
transition-metal monoxides possess the fcc structure
$Fm3m = \Gamma^f_cO^5_h$ (225) (in parentheses the international
number), CoO occupies a special position in the magnetic phase: the
magnetic structures of MnO, FeO and NiO are known to be monoclinic
base-centered~\cite{rooksby,shull,rothI,cj}, but the magnetic
structure of CoO is not fully understood. Two different models of
antiferromagnetic CoO are discussed in the literature: first, the
non-collinear multi-spin-axis magnetic structure with tetragonal
symmetry proposed in 1964 by van Laar~\cite{vanlaar} and, second, a
collinear monoclinic structure similar to that of the other
monoxides. The models can hardly be distinguished by neutron
diffraction data or by reverse Monte Carlo refinements of these
data~\cite{timm}. They are considered as alternative structures or as
structures coexisting in antiferromagnetic CoO~\cite{tomiyasu}. The
non-collinear structure is suggested by the marked tetragonal
distortion of CoO accompanying the antiferromagnetic state, and a collinear
structure could be associated with the additional small monoclinic
deformation unambiguously detected in the antiferromagnetic phase of
CoO~\cite{jauch}.

The present paper reports theoretical evidence that the tetragonal
multi-spin-axis structure does exist in CoO and creates the tetragonal
distortion of the crystal. The additional small monoclinic deformation
is not connected with the magnetic order but is evidently a
``monoclinic-like'' deformation of the array of the oxygen atoms,
comparable to the rhombohedral-like distortion proposed to exist in
antiferromagnetic NiO~\cite{enio} and MnO~\cite{emno}.

In the following Section~\ref{sec:mgroup} the magnetic group of the
multi-spin-axis structure is determined. Just as in NiO and MnO, a
problem emerges in this context: a system invariant under the already
reported~\cite{vanlaar,timm} type IV Shubnikov magnetic group
$I_c4_1\!/\!acd$ given in Equation~\gl{eq:1} cannot possess
antiferromagnetic eigenstates.  As shown in
Section~\ref{sec:tetragonal}, for this reason the crystal undergoes a
marked tetragonal distortion reducing the symmetry initially defined
by the group $I_c4_1\!/\!acd$. As demonstrated in
Section~\ref{sec:monoclinic}, this tetragonal distortion evidently
produces a monoclinic-like deformation of the array of the oxygen
atoms in the deformed crystal.

In Section~\ref{sec:wannierf} the nonadiabatic Heisenberg model (NHM)
shall be applied to the electronic ground state of CoO.  In
Section~\ref{sec:paramagneticwf}, we will show that the electrons
of paramagnetic CoO may occupy an atomic-like state allowing the
system to be a Mott insulator. Having in Section~\ref{sec:mgroup}
determined the active magnetic group of antiferromagnetic CoO, we may
apply the NHM to the antiferromagnetic state of CoO, too. In
Section~\ref{sec:antiferromagneticwf}, we will determine the magnetic
band related to the magnetic group of the antiferromagnetic state. The
electrons may perform an atomic-like motion in the nonadiabatic system
stabilizing the tetragonal non-collinear magnetic structure. In
addition, this atomic-like motion allows the electron system to be
Mott insulating.

\section{Magnetic group of the antiferromagnetic state}
\label{sec:mgroup}

\begin{figure}[!]
  \centering

  \includegraphics[width=.49\textwidth]{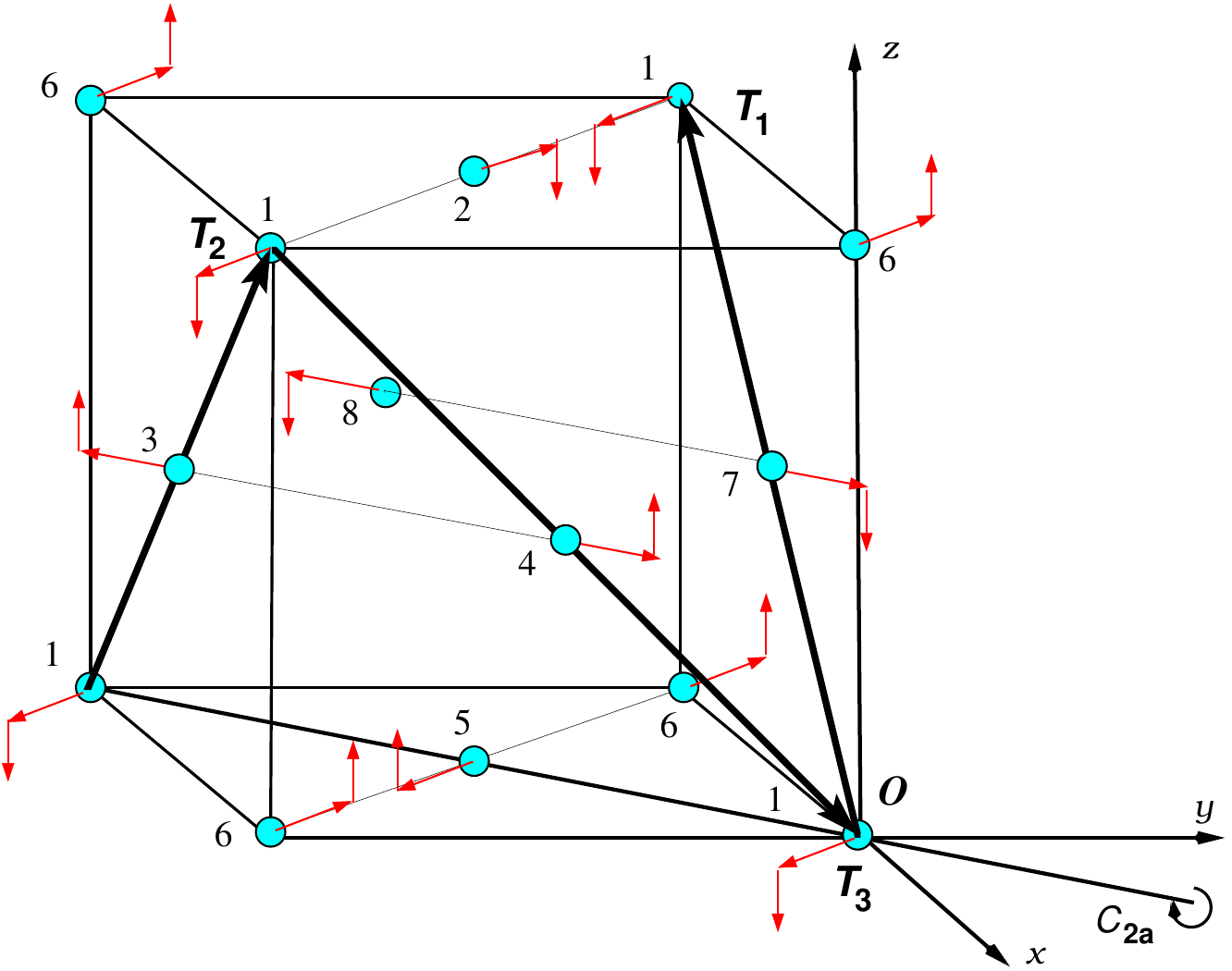}
  \includegraphics[width=.49\textwidth]{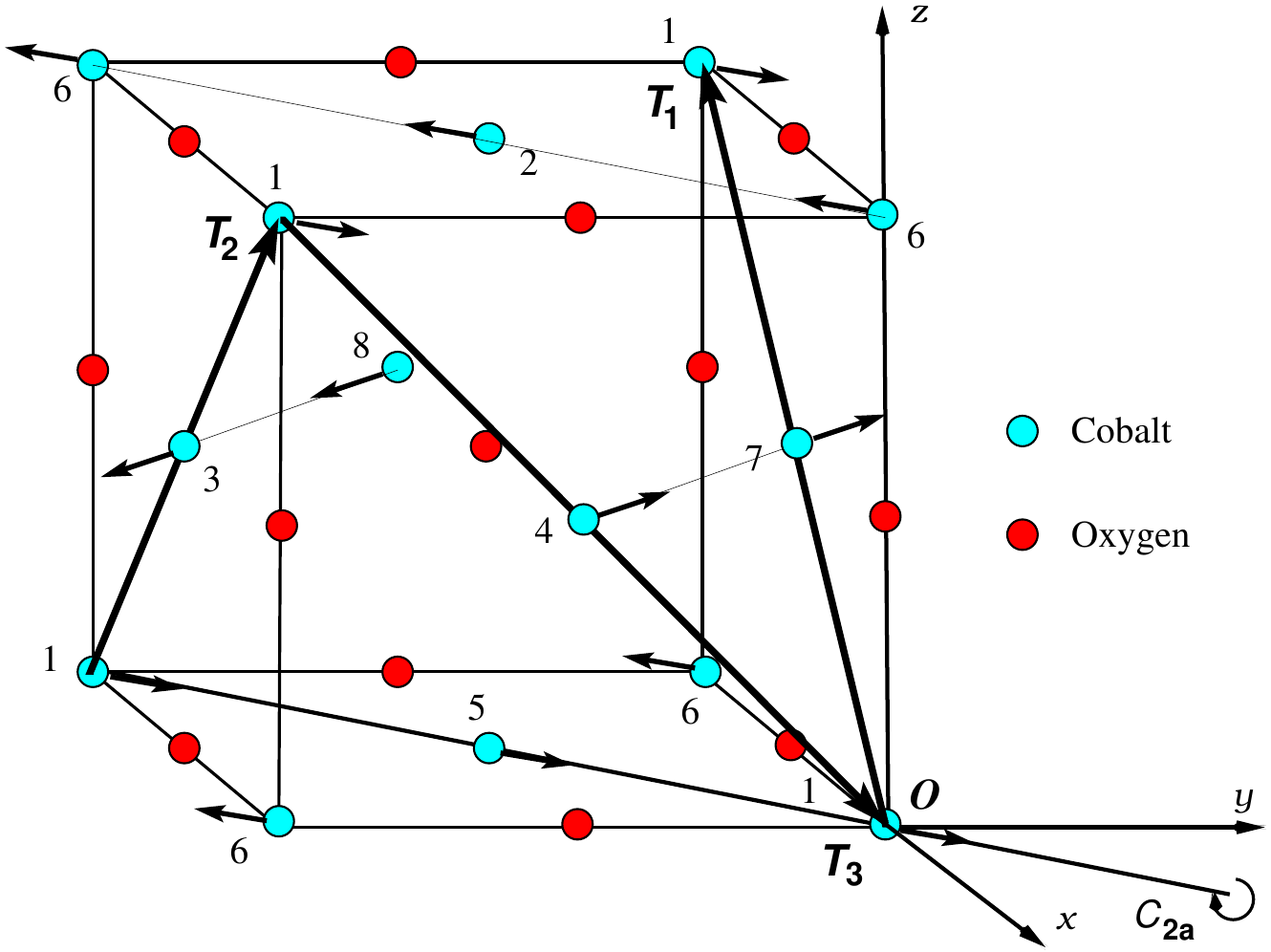}

  {\bf (a)} \hspace{.4\textwidth} {\bf (b)}
\caption{Antiferromagnetic CoO with the magnetic group $M_{110}$
  (Equation~\gl{eq:15}) based on the tetragonal body-centered Bravais
  lattice $\Gamma^v_q$. The eight Co atoms in the unit cell
  are (arbitrarily) numbered from 1 to 8. The three vectors $\bm T_i$
  are the basic translations of $\Gamma^v_q$. {\bf (a)} Magnetic
  structure of CoO invariant under $M_{110}$. The O atoms are not
  shown. The red arrows denote two components
  of the magnetic moments at the Co atoms. The components parallel to the
  plain (001) lie in $\pm(\bm T_1 + \bm T_3)$ or $\pm(\bm T_2 + \bm T_3)$
  direction and are all of the same size, the other components in $\pm z$ direction
  are also of the same size. The figure is consistent with Figure 3 of
  \cite{vanlaar}. {\bf (b)} Distortion of CoO invariant under
  $M_{110}$.
  The arrows denote the shifts of
  the Co atoms stabilizing the antiferromagnetic structure. The drawn
  shifts in $\pm(\bm T_1 + \bm T_3)$ or $\pm(\bm T_2 + \bm
  T_3)$ direction form the only distortion of the antiferromagnetic
  lattice invariant under $M_{110}$. Any
  shift of the O atoms, on the other hand, cannot happen 
  because it would destroy the
  symmetry of $M_{110}$.  
\label{fig:structures}
}
\end{figure}   
The multi-spin-axis structure of CoO~\cite{vanlaar} is indicated in
Figure~\ref{fig:structures} (a) by red arrows. It is invariant under the
space group $I4_1\!/\!acd$ (142)~\cite{vanlaar} and under the type IV
Shubnikov magnetic group~\cite{timm}:
\begin{equation}
  \label{eq:1}
  I_c4_1\!/\!acd =  I4_1\!/\!acd + \{K|{\textstyle \frac{1}{2}\frac{1}{2}}0 \}I4_1\!/\!acd, 
\end{equation}
where $K$ denotes the anti-unitary operator of time-inversion. The
unitary subgroup $I4_1\!/\!acd$ has the tetragonal body-centered
Bravais lattice $\Gamma^v_q$ and contains (besides the pure
translations) sixteen elements which are expressible as products of
the three generating elements
\begin{equation}
  \label{eq:9}
  \textstyle  \{C^+_{4z}|\frac{1}{2}00\},
  \{\sigma_{da}|\frac{1}{2}\frac{1}{2}0\}, \text{ and } \{I|000\}.
\end{equation}
We write the symmetry operations $\{R|\alpha\beta\gamma\}$ in the
Seitz notation: $R$ is a point group operation written with respect to
the $x$, $y$, and $z$ axes in Figure~\ref{fig:structures} and
$\alpha\bm T_1 + \beta\bm T_2 + \gamma\bm T_3$ the subsequent
translation~\cite{bc}, where the basis translations $\bm T_i$ are also
defined in Figure~\ref{fig:structures}. We write the point group
operations $R$ as defined in Section 1.4 of \cite{bc}, here $R = E$
stands for the identity operation, $R = C^+_{4z}$ for the
anti-clockwise rotation through $90^{\circ}$ about the $z$ axis,
$R = I$ for the inversion, $R = C_{2a}$ for the rotation through $180^{\circ}$
about the $a$ axis as indicated in Figure~\ref{fig:structures}, and
$R = \sigma_{da}$ for the reflection $IC_{2a}$.  Since the magnetic
structure in Figure~\ref{fig:structures} (a) is invariant under the
three generating elements~\gl{eq:9} and under the anti-unitary element
$\{K|{\textstyle \frac{1}{2}\frac{1}{2}}0 \}$, it is invariant under
the complete magnetic group~\gl{eq:1}.

In what follows, the magnetic group $I_c4_1\!/\!acd$ is referred to as $M_{142}$
because the unitary subgroup $I4_1\!/\!acd$ bears the international number 142,
\begin{equation}
 \label{eq:31}
 M_{142} = I4_1\!/\!acd + \{K|{\textstyle \frac{1}{2}\frac{1}{2}}0 \}I4_1\!/\!acd. 
\end{equation}

Table~\ref{tab:rep_142} lists the character tables of the single
valued irreducible representations of $I4_1\!/\!acd$. In particular, the
entry (a) below $\{K|000\}$ means that all the one-dimensional
representations of $I4_1\!/\!acd$ follow Case (a) with respect to the gray
magnetic group $I4_1\!/\!acd + \{K|000\}I4_1\!/\!acd$. The Case (a) is
defined by Equation (7.3.45) of \cite{bc}, in the present case the
one-dimensional representations follow Case (a) because they are real.
According to Condition 1 in~\cite{enio}, the antiferromagnetic
state cannot be an eigenstate of a system invariant under $M_{142}$
because $I4_1\!/\!acd$ does not possess at least one one-dimensional
representation following Case (c) (as defined by Equation\ (7.3.47)
of~\cite{bc}). 

These group-theoretical findings change drastically when we
assume that the subgroup $I4_1\!/\!cd$ (110) of $I4_1\!/\!acd$ (142) is
the unitary part of the magnetic group of antiferromagnetic
CoO. $I4_1\!/\!cd$ contains one-half of the sixteen elements of
$I4_1\!/\!acd$ (142) and may be defined by the generating elements
\begin{equation}
  \label{eq:4}
    \textstyle  \{C^+_{4z}|\frac{1}{2}00\} \text{ and }
  \{\sigma_{da}|\frac{1}{2}\frac{1}{2}0\}, 
\end{equation}
see Table 3.7 of~\cite{bc}. However, the origin used in the Tables
of~\cite{bc} for the space group $I4_1\!/\!cd$ (110) is different from
the origin used in the present paper (and marked in
Figure~\ref{fig:structures} by $O$). Thus, all the symmetry operations
$S_{BC}$ used in~\cite{bc} for the group $I4_1\!/\!cd$ (110) must be
transformed by $\{E|0\frac{1}{4}\frac{3}{4}\}$ into the symmetry
operations $S_{paper}$ used in this paper,
\begin{equation}
  \label{eq:11}
 \textstyle S_{paper} = \{E|0\frac{1}{4}\frac{3}{4}\}S_{BC}\{E|0\frac{1}{4}\frac{3}{4}\}^{-1}.
\end{equation}

Just as the group $I4_1\!/\!acd$, the group
$I4_1\!/\!cd$ has the tetragonal body-centered Bravais lattice
$\Gamma^v_q$. From $I4_1\!/\!cd$ we may derive two magnetic groups,
the type IV Shubnikov group
\begin{equation}
  \label{eq:5}
  \textstyle M_1 =  I4_1\!/\!cd + \{K|\frac{1}{2}\frac{1}{2}0\}I4_1\!/\!cd
\end{equation}
(with a black and white Bravais lattice),
and the type III Shubnikov group
\begin{equation}
  \label{eq:6}
  \textstyle M_2 =  I4_1\!/\!cd + \{KI|\frac{1}{2}\frac{1}{2}0\}I4_1\!/\!cd
\end{equation}
(with an ordinary Bravais lattice),
both leaving invariant the magnetic structure. The one-dimensional
representations at point $Z$ in Table~\ref{tab:rep_110} now follow both
requirements (i) and (ii) of Condition 1 of~\cite{enio} for the
magnetic group $M_2$, but no one-dimensional representation of
$I4_1\!/\!cd$ meets both requirements (i) and (ii) for $M_1$.

Thus, the type III Shubnikov group $M_2$ is the magnetic group of
antiferromagnetic CoO because it allows the system to possess
antiferromagnetic eigenstates. In what follows, we refer $M_2$ to as
\begin{equation}
  \label{eq:10}
  \textstyle M_{110} = I4_1\!/\!cd + \{KI|\frac{1}{2}\frac{1}{2}0\}I4_1\!/\!cd.
\end{equation}

\section{Tetragonal distortion of antiferromagnetic CoO} 
\label{sec:tetragonal}
In the tetragonally distorted crystal the eight Co and eight O atoms
in the unit cell are located at the positions
\begin{equation}
\label{eq:7}
\begin{tabular}{cccc}
Co($000$) & Co($\frac{1}{2}1\frac{1}{2}$) & Co($1\frac{1}{2}1$) &
Co($11\frac{1}{2}$) \\  Co($0\frac{1}{2}\frac{1}{2}$) &
Co($\frac{1}{2}\frac{1}{2}0$) & Co($\frac{3}{2}11$) &
Co($\frac{1}{2}\frac{1}{2}\frac{1}{2}$)\\
\end{tabular}
\end{equation}
and
\begin{equation}
\label{eq:8}
\begin{tabular}{cccc}
 O($\frac{1}{4}\frac{1}{4}0$) & O($\frac{3}{4}\frac{5}{4}\frac{1}{2}$)
 & O($\frac{5}{4}\frac{3}{4}1$) &
 O($\frac{1}{4}\frac{1}{4}\frac{1}{2}$) \\
 O($\frac{3}{4}\frac{3}{4}0$) & O($\frac{1}{4}\frac{3}{4}\frac{1}{2}$)
 & O($\frac{3}{4}\frac{5}{4}1$) &
 O($\frac{3}{4}\frac{3}{4}\frac{3}{2}$),\\ 
\end{tabular}
\end{equation}
respectively, in the coordinate system defined by the basic
translations $\bm T_1$, $\bm T_2$, and $\bm T_3$ of $\Gamma^v_q$ given
in Figure~\ref{fig:structures}. The positions~\gl{eq:7} and~\gl{eq:8}
are the Wyckoff positions 16c and 16e (with x=0) of the space group
$I4_1\!/\!acd$ (142)~\cite{vanlaar}. It should be noted that the unit
cell of $\Gamma^v_q$ does not contain 16 Co atoms and 16 O atoms as is
often assumed (and is suggested by the notations 16c and 16e of the
Wyckoff positions). In the present paper, the point in the center of the
tetragonal prism is not (and must not be) an additional point within the
unit cell, but is a lattice point connected by translation symmetry
with the other lattice points. Thus, there are 8 Co and 8 O atoms in
the unit cell.  

The tetragonal magnetic structure produces initially a distortion
invariant under the magnetic group $M_{142}$ in antiferromagnetic CoO.
Since a system invariant under $M_{142}$ does not possess
antiferromagnetic eigenstates, the crystal must be additionally
distorted in such a way that the electronic Hamiltonian still commutes
with the symmetry operations of $M_{110}$, but does not commute with
the symmetry operations of ${M_{142} - M_{110}}$, cf. Section 3
of~\cite{enio}.  The only distortion bringing the desired effect is a
shift of the Co atoms in $\pm(\bm T_1 + \bm T_3)$ and
$\pm(\bm T_2 + \bm T_3)$ direction from their positions~\gl{eq:7}, as
indicated by the arrows in Figure~\ref{fig:structures} (b). In fact,
the magnetic structure together with the shifts in
Figure~\ref{fig:structures} (b) is invariant under the two generating
elements~\gl{eq:4} and under the anti-unitary element
$\{KI|\frac{1}{2}\frac{1}{2}0\}$ defining $M_{110}$, but the
shifts are not invariant under the inversion $\{I|000\}$ being a
generating element of $M_{142}$.  The oxygen atoms are not shifted
from their positions~\gl{eq:8} because any shift of the oxygen
atoms would be incompatible with the magnetic group $M_{110}$.

This result may be understood by inspection of
Figure~\ref{fig:structures}, but also in terms of Wyckoff positions:
the type III Shubnikov group $M_{110}$ may be written in the
form~\cite{bc}
\begin{equation}
  \label{eq:12}
  \textstyle M_{110} = I4_1\!/\!cd + \{K|000\}\Big(G - I4_1\!/\!cd\Big)
\end{equation}
where
\begin{equation}
  \label{eq:13}
 \textstyle  G = I4_1\!/\!cd + \{I|\frac{1}{2}\frac{1}{2}0\}I4_1\!/\!cd
\end{equation}
is an ordinary (unitary) space group.  Equation~\gl{eq:12} shows that
the Wyckoff positions of $M_{110}$ are equal to the Wyckoff positions
of $G$, since the operator $\{K|000\}$ of time inversion does not
influence the atomic positions.  $G$ is equivalent to the group
$I4_1\!/\!acd$ (142) because we obtain the elements of $G$ when we
transform the elements of $I4_1\!/\!acd$ by the translation
$\{E|\frac{1}{4}\frac{1}{4}0\}$. This statement may be demonstrated by
means of the generating elements
\begin{eqnarray}
  \label{eq:14}
 \textstyle  \{C^+_{4z}|\frac{1}{2}00\} & = & \textstyle \{E|\frac{1}{4}\frac{1}{4}0\}\{C^+_{4z}|\frac{1}{2}00\}\{E|\frac{1}{4}\frac{1}{4}0\}^{-1},\nonumber\\
 \textstyle  \{\sigma_{da}|\frac{1}{2}\frac{1}{2}0\} & = & \textstyle \{E|\frac{1}{4}\frac{1}{4}0\}\{\sigma_{da}|\frac{1}{2}\frac{1}{2}0\}\{E|\frac{1}{4}\frac{1}{4}0\}^{-1},\\
 \textstyle  \{I|\frac{1}{2}\frac{1}{2}0\} & = & \textstyle \{E|\frac{1}{4}\frac{1}{4}0\}\{I|000\}\{E|\frac{1}{4}\frac{1}{4}0\}^{-1},\nonumber
\end{eqnarray}
see Equation 1.5.12 and Table 3.2 of~\cite{bc}.
Consequently, we receive the origin of $G$ by shifting the origin of 
$I4_1\!/\!acd$ (142) about
\begin{equation}
  \label{eq:15}
 \textstyle \bm t_0 = \frac{1}{4}\bm T_1 + \frac{1}{4}\bm T_2,
\end{equation}
that means, by shifting the origin from a Co atom to an O atom,
see Figure~\ref{fig:structures}. Thus, the Wyckoff positions are
interchanged: In the group $G$ and, hence, in the magnetic group
$M_{110}$, the Co atoms lie on the position $16e$ of the group
$I4_1\!/\!acd$ (142), and the O atoms on the position $16c$.  The
position $16e$ contains a parameter (usually $x$) describing the
shifts of the Co atoms indicated in Figure~\ref{fig:structures}
(b), the O atoms on position $16c$ are fixed. This fact plays an
essential role in the following Section~\ref{sec:monoclinic}.

\section{Monoclinic-like distortion}
\label{sec:monoclinic}
As is the case with antiferromagnetic NiO and MnO, also
antiferromagnetic CoO is clearly deformed by the magnetic structure
but, additionally, by a slight distortion seemingly incompatible with
the magnetic structure.  In NiO~\cite{enio} and MnO~\cite{emno}, this
additional distortion is evidently produced by the oxygen atoms which
form an rhombohedral-{\em like} array within the monoclinic magnetic
group $M_9$ of antiferromagnetic NiO and MnO. This distortion was called
``inner distortion'' of the magnetic group $M_9$ because the symmetry
of $M_9$ is not disturbed (and must not be disturbed) by the
rhombohedral-like distortion. In CoO, on the other hand, the magnetic
group $M_{110}$ is tetragonal and the additional distortion proved
experimentally to be monoclinic~\cite{jauch}. This can be understood
as follows:

Figure~\ref{fig:structures} shows the distorted antiferromagnetic
structure of CoO with the magnetic group $M_{110}$.  The arrows in
Figure~\ref{fig:structures} {\bf (b)} specify the shifts of the
cobalt atoms from their positions in Equation~\gl{eq:7}. These
shifts stabilize, on the one hand, the antiferromagnetic
structure (Section~\ref{sec:mgroup}) and produce, on the other hand, a
strong tetragonal distortion of the crystal. Within this distortion,
adjacent Co atoms are dislocated either in the same or in different
directions and, consequently, their distance and, hence, their mutual
attraction varies around its value in the paramagnetic state.
Consequently, the crystal is tetragonally distorted in such a way that
the basic translations $\bm T_1$, $\bm T_2$, and $\bm T_3$ of
$\Gamma^v_q$ are no longer embedded in the cubic lattice since this
lattice no longer exists.

\begin{figure}[!]
  \centering

  \includegraphics[width=0.5\textwidth]{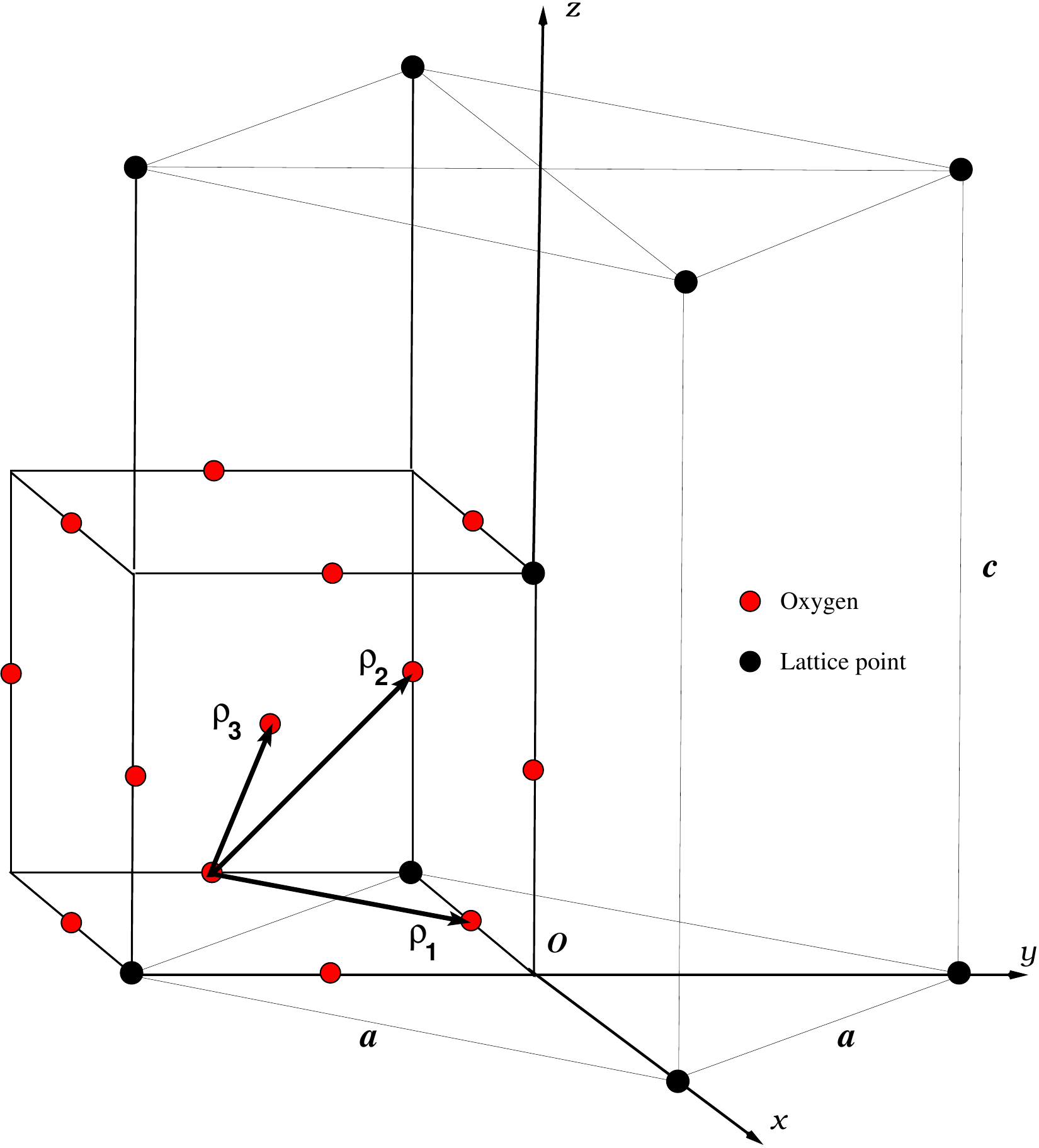}

  \caption{Oxygen atoms in the body-centered tetragonal Bravais
    lattice $\Gamma^v_q$. The Co atoms are not shown. 
\label{fig:oxygen}
}
\end{figure}   

The oxygen atoms, on the other hand, are not shifted from their
positions in the tetragonal body-centered lattice, that means, from
their position given in the list~\gl{eq:8}. Thus, their mutual
distances are essentially the same as in the paramagnetic phase. Just
as in NiO and MnO, they form a periodic array within the lattice of
the Co atoms. We assume again that this array forms a Bravais lattice
which does not contain eight O atoms but only one O atom in the unit
cell.  The array of the oxygen atoms is spanned by the vectors
$\bm \rho_1$, $\bm \rho_2$, and $\bm \rho_3$ in
Figure~\ref{fig:oxygen}.  Though these vectors are symmetry operations
in the paramagnetic lattice, they are no translation operators in the
tetragonal distorted crystal. Nevertheless, they give an approximate
picture of the array of the oxygen atoms. In antiferromagnetic NiO and
MnO, the vectors $\bm \rho_i$ define an rhombohedral-like array of the
oxygen atoms because they form a trigonal basis within the monoclinic
crystals of NiO and MnO, see Section 4 of~\cite{enio}.  In CoO, on the
other hand, this basis is not trigonal because the vectors
$\bm \rho_i$ have different lengths: the length of $\bm \rho_1$ is
different from the lengths of $\bm \rho_2$ and $\bm \rho_3$ in the
tetragonally deformed crystal,
\begin{equation}
  \label{eq:3}
  |\bm \rho_1|  \neq  |\bm \rho_2|  =  |\bm \rho_3|,
\end{equation}
if $c \neq \sqrt{2} a$ (i.e., if the crystal is no longer cubic), see
Figure~\ref{fig:oxygen}. Thus, the vectors $\bm \rho_i$ define a
monoclinic base-centered array of the oxygen atoms, see Table 3.1
of~\cite{bc}.  As stated above, this array is not exactly monoclinic
because the vectors $\bm \rho_i$ are no translational symmetry
operations in the distorted crystal. Therefore I call it a
``monoclinic-like'' distortion of antiferromagnetic CoO being an
``inner distortion'' of the magnetic group $M_{110}$. This will say
that the monoclinic-like distortion of the oxygen atoms is not
connected with any change or modification of the tetragonal symmetry
of the antiferromagnetic state which remains invariant under the
symmetry operations of the magnetic group $M_{110}$.

\section{Application of the Nonadiabatic Heisenberg Model}
\label{sec:wannierf}
The NHM defines in narrow, partly filled electronic energy bands a
strongly correlated nonadiabatic atomic-like motion. The nonadiabatic
localized states defining this atomic-like motion are represented by
symmetry-adapted and optimally localized Wannier
functions~\cite{enhm}. In the following
Subsection~\ref{sec:paramagneticwf} we will show that the band
structure of paramagnetic CoO contains an ``insulating
band''~\cite{eniopara} whose Bloch functions can be unitarily
transformed into symmetry-adapted and optimally localized Wannier
functions including all the electrons at the Fermi level. In the
second Subsection~\ref{sec:antiferromagneticwf} we will show that the
band structure of antiferromagnetic CoO encloses two ``magnetic
bands''~\cite{enio} related to the magnetic group $M_{110}$ of the
antiferromagnetic phase. The Bloch functions of these two bands can be
unitarily transformed into optimally localized Wannier functions
symmetry-adapted to $M_{110}$. They are even ``magnetic super
bands''~\cite{enio} since all the electrons at the Fermi level belong
to these two magnetic band.

\subsection{Atomic-like Electrons in Paramagnetic CoO}
\label{sec:paramagneticwf}

 \begin{figure*}[h]
 \includegraphics[width=.95\textwidth,angle=0]{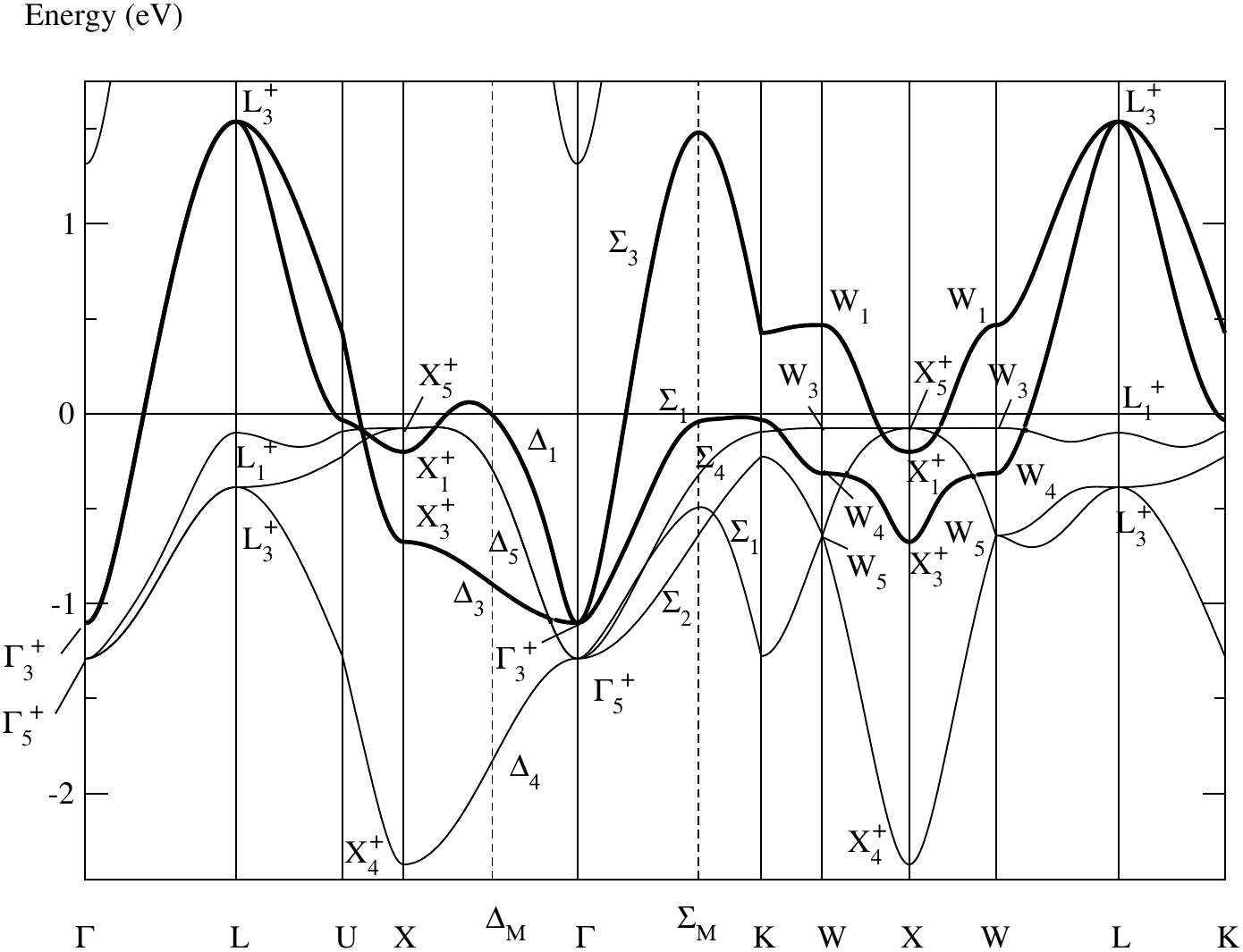}%
 \caption{Conventional band structure (Section 5 of~\cite{emno}) of
   paramagnetic fcc CoO as calculated by the FHI-aims program
   \cite{blum1,blum2} using the length $a = 4.260$ {\AA} of the
   paramagnetic unit cell as given in~\cite{bredow}. The symmetry
   labels as defined in 
Table A1 of~\cite{enio} are determined by the
author (as described in Section 2 of~\cite{enio}), the symmetry on the lines $\Sigma$ and $\Delta$ are defined by
Table~\ref{tab:comp_rel_225_110} {\bf (a)}. The notations of the
points and lines of symmetry in the Brillouin zone
for $\Gamma^f_c$ follow Figure 3.14 of Ref.~\cite{bc}. The insulating
band is highlighted by the bold line.
 \label{fig:bandstr225}
}
 \end{figure*}



Figure~\ref{fig:bandstr225} shows  the band structure of paramagnetic
CoO. The band indicated by the bold lines is characterized by Bloch
functions with the symmetry:
\begin{equation}
 \label{eq:2}
 \Gamma^+_3,\ L^+_3,\  X^+_1 + X^+_3,\ W_1 + W_4 
\end{equation}
of band 5 listed Table~\ref{tab:wf_225} (a). Thus, the Bloch functions
of this band can be unitarily transformed into optimally localized
Wannier functions adapted to the cubic fcc symmetry and centered at
the Co atoms. On the other hand, the Bloch functions cannot be
unitarily transformed in such a way that the Wannier functions are
centered at the O atoms since the symmetry of the Bloch functions at
point $L$ does not coincide with the symmetry of band 5 in
Table~\ref{tab:wf_225} (b). The Bloch functions~\gl{eq:2} define a
band with two branches yielding two degenerate Wannier functions with
$\Gamma^+_3$ symmetry at each Co atom. There symmetry may be called
``$d$-like'' since the energy band~\gl{eq:2} originates entirely from
a $d$ orbital of Co, see {Table 2.7} of~\cite{bc}. If this $d$-like
band is half-filled, it defines an atomic-like motion qualifying
paramagnetic CoO to be a Mott insulator because it consists of all the
branches crossing the Fermi level. In the paramagnetic system the
atomic-like motion occurs solely between the Co atoms.

\subsection{Atomic-like Electrons in Antiferromagnetic CoO}
\label{sec:antiferromagneticwf}


 \begin{figure*}[!]
\centering
\includegraphics[width=.75\textwidth,angle=0]{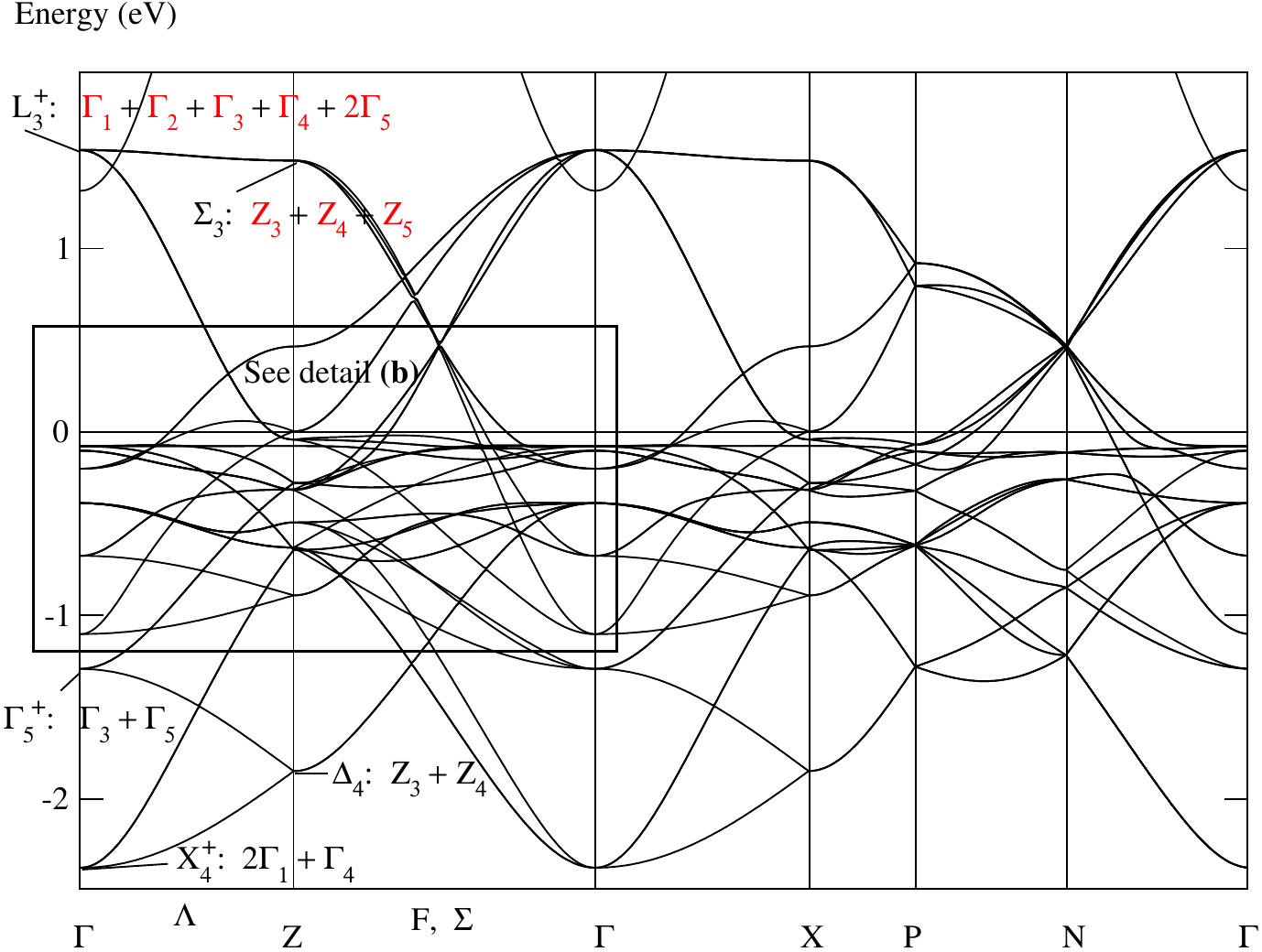}

{\bf (a)}
\ \\
\ \\
\ \\

\includegraphics[width=.75\textwidth,angle=0]{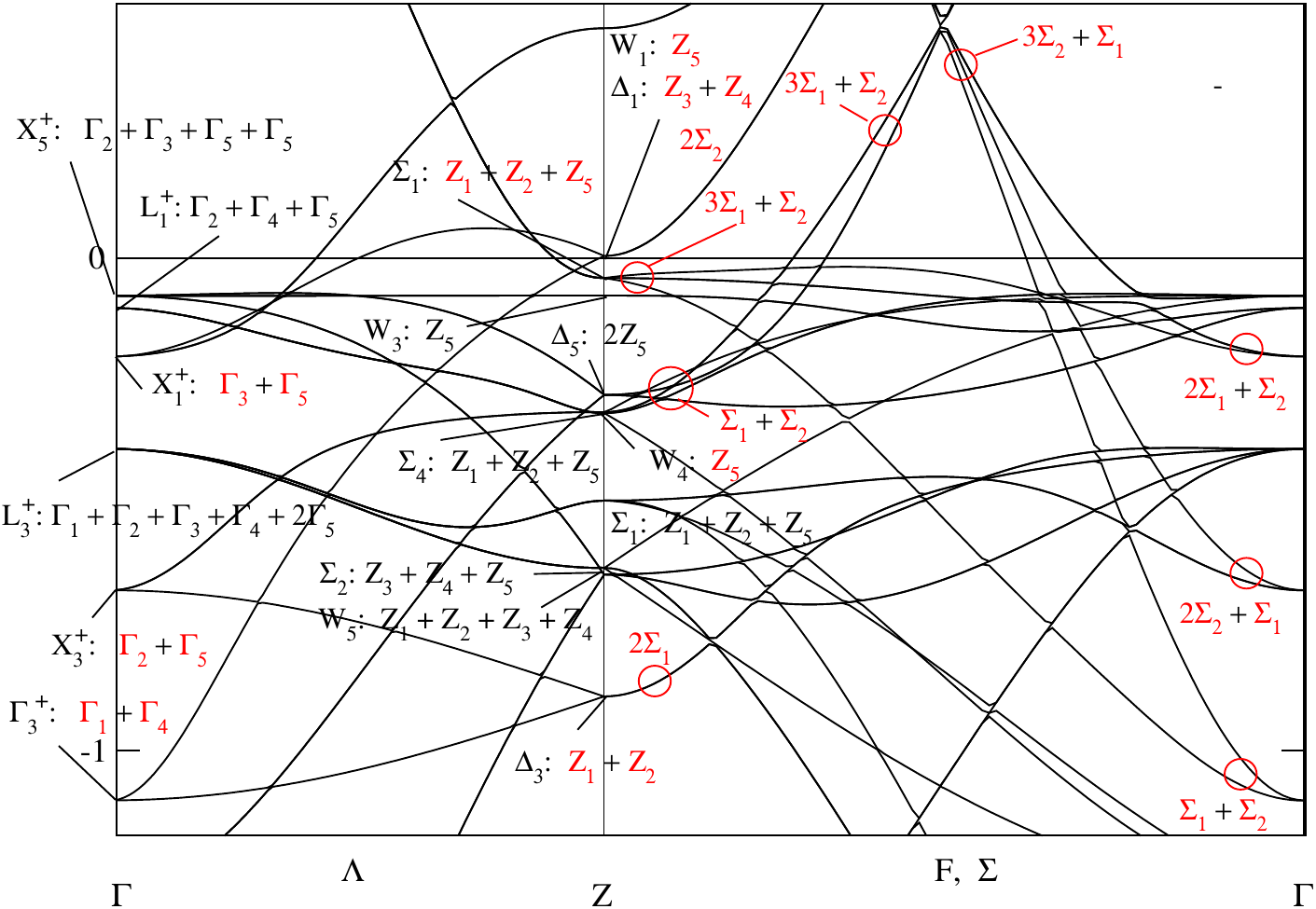}

{\bf (b)}
 \caption{ The band structure of CoO given in
   Figure~\ref{fig:bandstr225} folded into the Brillouin zone for the
   tetragonal body-centered lattice $\Gamma^v_q$ of the
   magnetic group $M_{110}$. As in Figure~\ref{fig:bandstr225}, the
   bands are calculated by the FHI-aims program \cite{blum1,blum2}.
   At points $\Gamma$ and $Z$, first the symmetry of the Bloch
   functions at the equivalent point in the fcc Brillouin zone is
   given. Then, after the colon, the symmetry of these Bloch functions
   in the tetragonal body-centered Brillouin zone is specified as
   given in Table~\ref{tab:falten225_110}.
   On the line $\Sigma$,$F$, the symmetry labels $\Sigma_1$ and
   $\Sigma_2$ are determined by Table~\ref{tab:comp_rel_225_110} {(b)}. 
   The Bloch functions highlighted in red form two
   magnetic bands (i.e., twice the band in Table~\ref{tab:wf_110})
   consisting
   of sixteen branches assigned to
   the eight cobalt and eight oxygen atoms. The
   notations of the points and lines of symmetry follow Figure~3.10 (b)~of
   Ref.~\cite{bc}.
\label{fig:bandstr110}
}
 \end{figure*}

Table~\ref{tab:wf_110} lists the only magnetic band related to the
magnetic group $M_{110}$ of antiferromagnetic CoO.
Figure~\ref{fig:bandstr110} shows the band structure of paramagnetic
CoO in Figure~\ref{fig:bandstr225} as folded into the Brillouin zone
$\Gamma^v_q$ of the tetragonal body-centered magnetic structure. The
band highlighted in Figure~\ref{fig:bandstr225} by the bold line (the
insulating band) becomes in Figure~\ref{fig:bandstr110} twice the
magnetic band in Table~\ref{tab:wf_110}. The Bloch functions defining
the sixteen branches of these two magnetic bands are highlighted in
red. They can be unitarily transformed into optionally three types of
optimally localized Wannier functions symmetry-adapted to $M_{110}$:
Either (i) the Wannier functions are centered twice at the eight Co
atoms (that means two Wannier functions are at each Co atom), or (ii)
they are centered twice at the eight O atoms; or (iii) they are
centered at both the eight Co atoms and at the eight O atoms.

The atomic-like motion defined by the first and the second possibility
(i) and (ii) clearly has a higher Coulomb energy than the atomic-like
motion belonging to the third possibility (iii) because the Coulomb
repulsion between two electrons occupying localized states at the same
atom is greater than the repulsion of two electrons occupying
localized states at different (adjacent) atoms. Thus, the last case
(iii) defines the energetically most favorable nonadiabatic
atomic-like motion which stabilizes the antiferromagnetic structure
with the magnetic group $M_{110}$~\cite{enio,ea}. Moreover, the two
magnetic bands are magnetic super bands since they comprise all the
branches crossing the Fermi level and, thus, qualify antiferromagnetic
CoO to be a Mott insulator~\cite{enio}.  While in the paramagnetic
system the atomic-like motion occurs solely between the Co atoms
(Section~\ref{sec:paramagneticwf}), it comprises in the
antiferromagnetic phase both, the Co and O atoms.

It should be noted that we can define in the band structure of
antiferromagnetic CoO other narrow bands differing (slightly) from the
bands highlighted in red, but coinciding also with the symmetry of the
band in Table~\ref{tab:wf_110}. They form likewise magnetic bands if
they are half-filled. It is not easy to say which bands represent best
the nonadiabatic atomic-like motion in antiferromagnetic CoO. In any
case, the two highlighted bands are a first approach to describing the
atomic-like motion because, in this case, the localized states in
antiferromagnetic CoO most resemble the localized states in the
paramagnetic phase. At this stage, we realize that two magnetic bands
exist, but we do not know the exact combination of their branches.

\section{Results}
\label{sec:results}
Under the assumption that the non-collinear multi-spin-axis
magnetic structure as proposed by van Laar is realized
in antiferromagnetic CoO, essential features of CoO can be understood:
\begin{itemize}
\item The non-collinear magnetic structure is invariant under the
  magnetic group $M_{142} = I_c4_1\!/\!acd$
  (Equation~\ref{eq:31}). However, an electronic Hamiltonian invariant
  under time inversion cannot possess magnetic eigenstates
  (Section~\ref{sec:mgroup}) if it commutes with all the symmetry
  operations of $M_{142}$.  For this reason, the crystal is markedly
  tetragonally deformed (see Section~\ref{sec:tetragonal} and
  Figure~\ref{fig:structures}) in order that the group $M_{110}$ in
  Equation~\ref{eq:10} becomes the magnetic group of antiferromagnetic
  CoO.
\item This tetragonal deformation produces a monoclinic-like
  deformation of the array of the oxygen atoms comparable with the
  rhombohedral-like deformation proposed in NiO and MnO
  (Section~\ref{sec:monoclinic}). This monoclinic-like deformation
  does not influence the magnetic structure, and thus, does not
  modify the magnetic group $M_{110}$. Consequently, no inconsistency
  between the tetragonal and monoclinic-like distortion remains.
\item In the band structure of CoO exist two magnetic bands related to
  $M_{110}$ (Section~\ref{sec:antiferromagneticwf}). One of the two
  bands yields Wannier functions centered at the eight Co atoms, the
  other yields Wannier functions centered at the eight O atoms. In
  each case, the Wannier functions are optimally localized and adapted
  to the symmetry of the magnetic state. Thus, the electrons of this
  band may perform a nonadiabatic atomic-like motion lowering their
  nonadiabatic condensation energy $\Delta E$ (as defined in Equation
  (2.20) in~\cite{enhm}) by activating an exchange mechanism producing
  a magnetic structure with the magnetic group $M_{110}$
  (Section~\ref{sec:wannierf}).
\item The two magnetic bands are magnetic super bands because they
  comprise all the branches crossing the Fermi level. Thus, they
  qualify antiferromagnetic CoO to be a Mott insulator
  (Section~\ref{sec:antiferromagneticwf}). The nonadiabatic
  atomic-like motion involves both, the Co and the O atoms.
\end{itemize}
Independently of the special magnetic symmetry it was shown in
Section~\ref{sec:paramagneticwf} that
\begin{itemize}
\item there exists an insulating band~\cite{enio} in the paramagnetic
  band structure of CoO. The electrons of this band may perform a
  nonadiabatic atomic-like motion that qualifies CoO to be a Mott
  insulator also in the paramagnetic phase. The related optimally
  localized Wannier functions are adapted to the cubic fcc symmetry of
  paramagnetic CoO, are centered on the Co atoms, are twofold
  degenerate and have d-like $\Gamma^+_3$ symmetry. The atomic-like
  motion occurs exclusively between the Co atoms.
\end{itemize}
\section{Discussion}
\label{sec:discussion}
The results of the present paper as listed in the preceding
Section~\ref{sec:results} provide strong evidence that the tetragonal
non-collinear multi-spin-axis structure exists in CoO.  The
foundations of the presented theory are
\begin{enumerate}
\item 
the nonadiabatic Heisenberg model (NHM) published 20 years ago~\cite{enhm}, and
\item the group-theoretical theorem, stating that in a system
  invariant under time inversion, any magnetic eigenstate and its
  time-inverted state belong to a two-dimensional irreducible
  co-representation of the magnetic
  group~\cite{ea,ef,bamn2as2,theoriewf}.
\end{enumerate}
Unfortunately, the three axioms of the NHM still are relatively
unfamiliar though they are easy to recognize. Nevertheless, since
being introduced, the NHM worked effectively
in understanding superconductivity~\cite{josn,ebi,theoriewf},
magnetism and Mott insulation~\cite{enio,eniopara,emno,bamn2as2}.  The
group-theoretical theorem (ii) proved to be useful already in
understanding distortions accompanying a magnetic
state~\cite{ea,lafeaso1,enio,emno}.
\vspace{6pt}



\funding{This publication was supported by the Open Access Publishing Fund of the University of Stuttgart.}
\acknowledgments{I am much indebted to Guido Schmitz for his
  continuing support of my work.}

\conflictsofinterest{The author declares no conflict of interest.}
\abbreviations{The following abbreviations are used in this manuscript:\\

\noindent 
\begin{tabular}{@{}ll}
NHM & Nonadiabatic Heisenberg model\\
$E$ & Identity operation\\
$I$ & Inversion\\
$C^+_{4z}$ & anti-clockwise rotation through $90^{\circ}$ about the $z$ axis\\
$C_{2a}$ & Rotation through $180^{\circ}$ as indicated in Figure~\ref{fig:structures}\\
$\sigma_{da}$ & Reflection $IC_{2a}$\\
$K$ & anti-unitary operator of time inversion
\end{tabular}}

\appendix
\section{Group-theoretical tables}
\label{sec:tables}
This appendix provides Tables~\ref{tab:rep_142} --~\ref{tab:comp_rel_225_110}
along with notes to the tables.
\FloatBarrier
\begin{table}[!]
\caption{
 Character tables of the single valued irreducible representations of the
 tetragonal body-centered space group $I4_1\!/\!acd = \Gamma^v_q D^{20}_{4h}$ (142).
 \label{tab:rep_142}}
\centering
\begin{tabular}[t]{ccccccc}
\multicolumn{7}{c}{$\Gamma (000)$}\\
 &  & $$ & $$ & $\{C^-_{4z}|\frac{1}{2}\frac{1}{2}\frac{1}{2}\}$ & $\{C_{2b}|0\frac{1}{2}\frac{1}{2}\}$ & $\{C_{2x}|00\frac{1}{2}\}$\\
 & $\{K|000\}$ & $\{E|000\}$ & $\{C_{2z}|\frac{1}{2}0\frac{1}{2}\}$ & $\{C^+_{4z}|\frac{1}{2}00\}$ & $\{C_{2a}|\frac{1}{2}\frac{1}{2}0\}$ & $\{C_{2y}|0\frac{1}{2}0\}$\\
\hline
$\Gamma^+_1$ & (a) & 1 & 1 & 1 & 1 & 1\\
$\Gamma^+_2$ & (a) & 1 & 1 & 1 & -1 & -1\\
$\Gamma^+_3$ & (a) & 1 & 1 & -1 & 1 & -1\\
$\Gamma^+_4$ & (a) & 1 & 1 & -1 & -1 & 1\\
$\Gamma^+_5$ & (a) & 2 & -2 & 0 & 0 & 0\\
$\Gamma^-_1$ & (a) & 1 & 1 & 1 & 1 & 1\\
$\Gamma^-_2$ & (a) & 1 & 1 & 1 & -1 & -1\\
$\Gamma^-_3$ & (a) & 1 & 1 & -1 & 1 & -1\\
$\Gamma^-_4$ & (a) & 1 & 1 & -1 & -1 & 1\\
$\Gamma^-_5$ & (a) & 2 & -2 & 0 & 0 & 0\\
\hline\\
\end{tabular}\hspace{1cm}
\begin{tabular}[t]{cccccc}
\multicolumn{6}{c}{$\Gamma (000)$\qquad $(continued)$}\\
 & $$ & $$ & $\{S^+_{4z}|\frac{1}{2}\frac{1}{2}\frac{1}{2}\}$ & $\{\sigma_{db}|0\frac{1}{2}\frac{1}{2}\}$ & $\{\sigma_x|00\frac{1}{2}\}$\\
 & $\{I|000\}$ & $\{\sigma_z|\frac{1}{2}0\frac{1}{2}\}$ & $\{S^-_{4z}|\frac{1}{2}00\}$ & $\{\sigma_{da}|\frac{1}{2}\frac{1}{2}0\}$ & $\{\sigma_y|0\frac{1}{2}0\}$\\
\hline
$\Gamma^+_1$ & 1 & 1 & 1 & 1 & 1\\
$\Gamma^+_2$ & 1 & 1 & 1 & -1 & -1\\
$\Gamma^+_3$ & 1 & 1 & -1 & 1 & -1\\
$\Gamma^+_4$ & 1 & 1 & -1 & -1 & 1\\
$\Gamma^+_5$ & 2 & -2 & 0 & 0 & 0\\
$\Gamma^-_1$ & -1 & -1 & -1 & -1 & -1\\
$\Gamma^-_2$ & -1 & -1 & -1 & 1 & 1\\
$\Gamma^-_3$ & -1 & -1 & 1 & -1 & 1\\
$\Gamma^-_4$ & -1 & -1 & 1 & 1 & -1\\
$\Gamma^-_5$ & -2 & 2 & 0 & 0 & 0\\
\hline\\
\end{tabular}\hspace{1cm}
\begin{tabular}[t]{cccccccc}
\multicolumn{8}{c}{$Z (\frac{1}{2}\frac{1}{2}\overline{\frac{1}{2}})$}\\
 & $$ & $$ & $$ & $$ & $\{C_{2a}|\frac{1}{2}\frac{1}{2}1\}$ & $\{C_{2a}|\frac{1}{2}\frac{1}{2}0\}$ & $\{I|000\}$\\
 & $\{E|000\}$ & $\{E|001\}$ & $\{C_{2z}|\frac{1}{2}0\frac{3}{2}\}$ & $\{C_{2z}|\frac{1}{2}0\frac{1}{2}\}$ & $\{C_{2b}|0\frac{1}{2}\frac{3}{2}\}$ & $\{C_{2b}|0\frac{1}{2}\frac{1}{2}\}$ & $\{I|001\}$\\
\hline
$Z_1$ & 2 & -2 & 2 & -2 & 2 & -2 & 0\\
$Z_2$ & 2 & -2 & -2 & 2 & 0 & 0 & 0\\
$Z_3$ & 2 & -2 & 2 & -2 & -2 & 2 & 0\\
$Z_4$ & 2 & -2 & -2 & 2 & 0 & 0 & 0\\
\hline\\
\end{tabular}\hspace{1cm}
\begin{tabular}[t]{cccccccc}
\multicolumn{8}{c}{$Z (\frac{1}{2}\frac{1}{2}\overline{\frac{1}{2}})$\qquad $(continued)$}\\
 & $$ & $$ & $$ & $\{C_{2y}|0\frac{3}{2}0\}$ & $\{C^+_{4z}|\frac{1}{2}01\}$ & $\{\sigma_y|0\frac{1}{2}0\}$ & $\{S^-_{4z}|\frac{1}{2}00\}$\\
 & $$ & $$ & $$ & $\{C_{2y}|0\frac{1}{2}0\}$ & $\{C^-_{4z}|\frac{1}{2}\frac{1}{2}\frac{1}{2}\}$ & $\{\sigma_y|0\frac{1}{2}1\}$ & $\{S^+_{4z}|\frac{1}{2}\frac{1}{2}\frac{1}{2}\}$\\
 & $\{\sigma_z|\frac{1}{2}0\frac{1}{2}\}$ & $\{\sigma_{db}|0\frac{1}{2}\frac{1}{2}\}$ & $\{\sigma_{db}|0\frac{1}{2}\frac{3}{2}\}$ & $\{C_{2x}|00\frac{3}{2}\}$ & $\{C^-_{4z}|\frac{1}{2}\frac{1}{2}\frac{3}{2}\}$ & $\{\sigma_x|00\frac{3}{2}\}$ & $\{S^+_{4z}|\frac{1}{2}\frac{1}{2}\frac{3}{2}\}$\\
 & $\{\sigma_z|\frac{1}{2}0\frac{3}{2}\}$ & $\{\sigma_{da}|\frac{1}{2}\frac{3}{2}0\}$ & $\{\sigma_{da}|\frac{1}{2}\frac{1}{2}0\}$ & $\{C_{2x}|00\frac{1}{2}\}$ & $\{C^+_{4z}|\frac{1}{2}00\}$ & $\{\sigma_x|00\frac{1}{2}\}$ & $\{S^-_{4z}|\frac{1}{2}10\}$\\
\hline
$Z_1$ & 0 & 0 & 0 & 0 & 0 & 0 & 0\\
$Z_2$ & 0 & 2 & -2 & 0 & 0 & 0 & 0\\
$Z_3$ & 0 & 0 & 0 & 0 & 0 & 0 & 0\\
$Z_4$ & 0 & -2 & 2 & 0 & 0 & 0 & 0\\
\hline\\
\end{tabular}\hspace{1cm}
\ \\
\begin{flushleft}
Notes to Table~\ref{tab:rep_142}:
\end{flushleft}
\begin{enumerate}
\item The notations of the points of symmetry follow Figure 3.10 (b)
of \cite{bc}.
\item Only the two points of symmetry $\Gamma$ and $Z$ invariant under
  the complete space group are listed.
\item The character tables are determined from Table 5.7 in~\protect\cite{bc}. 
\item $K$ denotes the operator of time inversion. The entry (a) is
 determined by Equation\ (7.3.51) of \cite{bc} and indicates
 that the related co-representations of the magnetic group
 $I4_1\!/\!acd + \{K|000\}I4_1\!/\!acd$ follow Case (a) as defined in Equation\ (7.3.45) of
 \cite{bc}.
\end{enumerate}
\end{table}

\begin{table}[!]
\caption{
 Character tables of the single valued irreducible representations of the
 tetragonal body-centered space group $I4_1\!/\!cd =
 \Gamma^v_qC^{12}_{4v}$\ \ (110).
 \label{tab:rep_110}}
\centering
\begin{tabular}[t]{ccccccccc}
\multicolumn{9}{c}{$\Gamma (000)$}\\
 &  &  &  & $$ & $$ & $\{C^-_{4z}|\frac{1}{2}\frac{1}{2}\frac{1}{2}\}$ & $\{\sigma_{da}|\frac{1}{2}\frac{1}{2}0\}$ & $\{\sigma_y|0\frac{1}{2}0\}$\\
 & $\{K|000\}$ & $\{KI|\frac{1}{2}\frac{1}{2}0\}$ & $\{K|\frac{1}{2}\frac{1}{2}0\}$ & $\{E|000\}$ & $\{C_{2z}|\frac{1}{2}0\frac{1}{2}\}$ & $\{C^+_{4z}|\frac{1}{2}00\}$ & $\{\sigma_{db}|0\frac{1}{2}\frac{1}{2}\}$ & $\{\sigma_x|00\frac{1}{2}\}$\\
\hline
$\Gamma_1$ & (a) & (a) & (a) & 1 & 1 & 1 & 1 & 1\\
$\Gamma_2$ & (a) & (a) & (a) & 1 & 1 & 1 & -1 & -1\\
$\Gamma_3$ & (a) & (a) & (a) & 1 & 1 & -1 & 1 & -1\\
$\Gamma_4$ & (a) & (a) & (a) & 1 & 1 & -1 & -1 & 1\\
$\Gamma_5$ & (a) & (a) & (a) & 2 & -2 & 0 & 0 & 0\\
\hline\\
\end{tabular}\hspace{1cm}
\begin{tabular}[t]{cccccccc}
\multicolumn{8}{c}{$Z (\frac{1}{2}\frac{1}{2}\overline{\frac{1}{2}})$}\\
 &  &  &  & $$ & $$ & $$ & $$\\
 & $\{K|000\}$ & $\{KI|\frac{1}{2}\frac{1}{2}0\}$ & $\{K|\frac{1}{2}\frac{1}{2}0\}$ & $\{E|000\}$ & $\{C_{2z}|\frac{1}{2}0\frac{1}{2}\}$ & $\{C_{2z}|\frac{1}{2}0\frac{3}{2}\}$ & $\{E|100\}$\\
\hline
$Z_1$ & (c) & (a) & (c) & 1 & 1 & -1 & -1\\
$Z_2$ & (c) & (a) & (c) & 1 & 1 & -1 & -1\\
$Z_3$ & (c) & (a) & (c) & 1 & 1 & -1 & -1\\
$Z_4$ & (c) & (a) & (c) & 1 & 1 & -1 & -1\\
$Z_5$ & (a) & (a) & (a) & 2 & -2 & 2 & -2\\
\hline\\
\end{tabular}\hspace{1cm}
\begin{tabular}[t]{ccccccc}
\multicolumn{7}{c}{$Z (\frac{1}{2}\frac{1}{2}\overline{\frac{1}{2}})$\qquad $(continued)$}\\
 & $\{\sigma_{db}|0\frac{1}{2}\frac{3}{2}\}$ & $\{\sigma_{da}|\frac{1}{2}\frac{1}{2}1\}$ & $\{C^-_{4z}|\frac{1}{2}\frac{1}{2}\frac{1}{2}\}$ & $\{C^+_{4z}|\frac{1}{2}01\}$ & $\{\sigma_x|00\frac{1}{2}\}$ & $\{\sigma_y|0\frac{1}{2}0\}$\\
 & $\{\sigma_{da}|\frac{1}{2}\frac{1}{2}0\}$ & $\{\sigma_{db}|0\frac{1}{2}\frac{1}{2}\}$ & $\{C^+_{4z}|\frac{1}{2}00\}$ & $\{C^-_{4z}|\frac{1}{2}\frac{1}{2}\frac{3}{2}\}$ & $\{\sigma_y|0\frac{1}{2}1\}$ & $\{\sigma_x|00\frac{3}{2}\}$\\
\hline
$Z_1$ & 1 & -1 & i & -i & i & -i\\
$Z_2$ & 1 & -1 & -i & i & -i & i\\
$Z_3$ & -1 & 1 & -i & i & i & -i\\
$Z_4$ & -1 & 1 & i & -i & -i & i\\
$Z_5$ & 0 & 0 & 0 & 0 & 0 & 0\\
\hline\\
\end{tabular}\hspace{1cm}
\begin{tabular}[t]{ccccc}
\multicolumn{5}{c}{$P (\frac{1}{4}\frac{1}{4}\frac{1}{4})$}\\
 & $$ & $$ & $$ & $$\\
 & $\{E|000\}$ & $\{E|010\}$ & $\{E|020\}$ & $\{E|030\}$\\
\hline
$P_1$ & 2 & -2i & -2 & 2i\\
\hline\\
\end{tabular}\hspace{1cm}
\begin{tabular}[t]{ccccccc}
\multicolumn{7}{c}{$P (\frac{1}{4}\frac{1}{4}\frac{1}{4})$\qquad $(continued)$}\\
 & $\{\sigma_y|0\frac{3}{2}1\}$ & $\{\sigma_y|1\frac{3}{2}1\}$ & $\{C_{2z}|\frac{1}{2}0\frac{1}{2}\}$ & $\{C_{2z}|\frac{1}{2}0\frac{3}{2}\}$ & $\{\sigma_x|00\frac{7}{2}\}$ & $\{\sigma_x|00\frac{1}{2}\}$\\
 & $\{\sigma_y|0\frac{1}{2}0\}$ & $\{\sigma_y|0\frac{1}{2}1\}$ & $\{C_{2z}|\frac{1}{2}1\frac{3}{2}\}$ & $\{C_{2z}|\frac{3}{2}1\frac{3}{2}\}$ & $\{\sigma_x|00\frac{3}{2}\}$ & $\{\sigma_x|01\frac{3}{2}\}$\\
\hline
$P_1$ & 0 & 0 & 0 & 0 & 0 & 0\\
\hline\\
\end{tabular}\hspace{1cm}
\begin{tabular}[t]{ccccc}
\multicolumn{5}{c}{$N (0\frac{1}{2}0)$}\\
 & $\{E|000\}$ & $\{\sigma_{db}|0\frac{1}{2}\frac{1}{2}\}$ & $\{E|010\}$ & $\{\sigma_{db}|0\frac{3}{2}\frac{1}{2}\}$\\
\hline
$N_1$ & 1 & i & -1 & -i\\
$N_2$ & 1 & -i & -1 & i\\
\hline\\
\end{tabular}\hspace{1cm}
\begin{tabular}[t]{cccccc}
\multicolumn{6}{c}{$X (00\frac{1}{2})$}\\
 & $$ & $$ & $\{\sigma_x|00\frac{3}{2}\}$ & $\{C_{2z}|\frac{1}{2}0\frac{1}{2}\}$ & $\{\sigma_y|0\frac{1}{2}1\}$\\
 & $\{E|000\}$ & $\{E|001\}$ & $\{\sigma_x|00\frac{1}{2}\}$ & $\{C_{2z}|\frac{1}{2}0\frac{3}{2}\}$ & $\{\sigma_y|0\frac{1}{2}0\}$\\
\hline
$X_1$ & 2 & -2 & 0 & 0 & 0\\
\hline\\
\end{tabular}\hspace{1cm}
\ \\
\begin{flushleft}
Notes to Table~\ref{tab:rep_110}:
\end{flushleft}
\begin{enumerate}
\item The notations of the points of symmetry follow Figure 3.10 (b) 
of ~\protect\cite{bc}.
\item The character tables are determined from Table 5.7 in~\protect\cite{bc}. 
\item $K$ denotes the operator of time inversion. The entries (a) and
 (c) are determined by Equation\ (7.3.51) of \cite{bc}. They
 indicate whether the related co-representations of the magnetic
 groups $I4_1\!/\!cd + \{K|000\}I4_1\!/\!cd$, $I4_1\!/\!cd + \{KI|\frac{1}{2}\frac{1}{2}0\}I4_1\!/\!cd$, and
 $I4_1\!/\!cd + \{K|\frac{1}{2}\frac{1}{2}0\}I4_1\!/\!cd$ follow Case (a) or Case (c) as defined in
 Equations~(7.3.45) and (7.3.47), respectively, of \cite{bc}.
\end{enumerate}
\end{table}


\begin{table}[!]
\caption{
  Excerpt of Table 1 of~\cite{eniopara} showing only
  Band 5 with optimally localized Wannier functions of $\Gamma^+_3$
  symmetry centered at the Co atoms (Table {\bf (a)}) and the O atoms (Table {\bf (b)}).  
\label{tab:wf_225}}
\begin{center}
\begin{tabular}[t]{cccccc}
{\bf (a)} & Co($000$) & $\Gamma$ & $X$ & $L$ & $W$\\
\hline
Band 5 & $\Gamma^+_3$ & $\Gamma^+_3$ & $X^+_1$ + $X^+_3$ & $L^+_3$ & $W_1$ + $W_4$\\
\hline
{\bf (b)} & O($\overline{\frac{1}{2}}\frac{1}{2}\frac{1}{2}$) &
$\Gamma$ & $X$ & $L$ & $W$\\
\hline
Band 5 & $\Gamma^+_3$ & $\Gamma^+_3$ & $X^+_1$ + $X^+_3$ & $L^-_3$ &
$W_1$ + $W_4$\\
\hline\\
\end{tabular}
\end{center}
\end{table}


\begin{table}[!]
\caption{
Symmetry labels of the Bloch functions at the points of symmetry in
the Brillouin zone for $I4_1\!/\!cd$ (110) of the only 
energy band with symmetry-adapted and optimally  
localized Wannier functions centered at the Co $\big($Table {\bf
  (a)}$\big)$ as well as O
$\big($Table {\bf (b)}$\big)$ atoms, respectively. 
\label{tab:wf_110}}
\centering
\begin{tabular}[t]{cccccccccc}
{\bf (a)}& {\bf Co}  & Co($000$) & Co($\frac{1}{2}1\frac{1}{2}$) & Co($1\frac{1}{2}1$) & Co($11\frac{1}{2}$) & Co($0\frac{1}{2}\frac{1}{2}$) & Co($\frac{1}{2}\frac{1}{2}0$) & Co($\frac{3}{2}11$) & Co($\frac{1}{2}\frac{1}{2}\frac{1}{2}$)\\
{\bf (b)}& {\bf O} & O($\frac{1}{4}\frac{1}{4}0$) & O($\frac{3}{4}\frac{5}{4}\frac{1}{2}$) & O($\frac{5}{4}\frac{3}{4}1$) & O($\frac{1}{4}\frac{1}{4}\frac{1}{2}$) & O($\frac{3}{4}\frac{3}{4}0$) & O($\frac{1}{4}\frac{3}{4}\frac{1}{2}$) & O($\frac{3}{4}\frac{5}{4}1$) & O($\frac{3}{4}\frac{3}{4}\frac{3}{2}$)\\
\hline
& & $\bm{d}_{1}$ & $\bm{d}_{1}$ & $\bm{d}_{1}$ & $\bm{d}_{1}$ & $\bm{d}_{1}$ & $\bm{d}_{1}$ & $\bm{d}_{1}$ & $\bm{d}_{1}$ \\
\hline\\
\end{tabular}

\begin{tabular}[t]{cccccccc}
\multicolumn{7}{c}{$(continued)$}\\
{\bf (a)} & {\bf (b)} &$\{KI|\frac{1}{2}\frac{1}{2}0\}$ & $\Gamma$ & $Z$ & $N$ & $X$ & $P$ \\
\hline
& & OK & $\Gamma_1$ + $\Gamma_2$ + $\Gamma_3$ + $\Gamma_4$ + 2$\Gamma_5$ &
$Z_1$ + $Z_2$ + $Z_3$ + $Z_4$ + 2$Z_5$ & 4$N_1$ + 4$N_2$ & 4$X_1$ & 4$P_1$\\
\hline\\
\end{tabular}
\ \\[1cm]
\begin{flushleft}
Notes to Table~\ref{tab:wf_110}
\end{flushleft}
\begin{enumerate}
\item The symmetry of the band related to the Co atoms in Table {\bf
    (a)} coincides fully with the symmetry of the band related to the
  O atoms in Table {\bf (b)}.
\item The notations of the points of symmetry in the Brillouin zone
  for $\Gamma^v_q$ follow Figure 3.10 (b) of~\cite{bc}
  and the symmetry notations of the Bloch functions are defined in
  Table~\ref{tab:rep_110}.
\item The bands are determined following Theorem 5
  of~\cite{theoriewf}.
\item The point groups $G_{0Co}$ and $G_{0O}$ of the
  positions~\cite{theoriewf} of the Co and the O atoms contain only
  the identity operation:
  \begin{equation}
    \label{eq:24}
    G_{0Co} = G_{0O} = \Big\{ \{E|\bm 0\} \Big\}.
    \end{equation}
    Thus, the Wannier functions at the atoms belong to the
    simple representation
\begin{center}
\begin{tabular}[t]{cc}
 & $\{E|\bm 0\}$\\
\hline
$\bm{d}_{1}$ & 1\\
\hline\\
\end{tabular}\hspace{1cm}
\end{center}
of $G_{0Co}$ and $G_{0O}$.
\item  The table defines a band consisting of eight branches with Bloch
  functions that can be unitarily transformed into Wannier functions
  being:
\begin{itemize}
\item as well localized as possible; 
\item centered at the eight Co atoms {\bf (a)} or
  at the eight O atoms {\bf (b)}; and
\item symmetry adapted to the unitary subgroup $I4_1\!/\!cd$ (110) of the
  magnetic group $M_{110}$. In the present case this means that the
  symmetry operations of $I4_1\!/\!cd$ effect an interchange of the eight
  Wannier functions in the unit cell, cf. note (xi) of Table A2
  in~\cite{enio}.
\item The entry ``OK'' indicates that the Wannier functions follow not
  only Theorem 5, but also Theorem 7 of
  \cite{theoriewf}. Consequently, they may be chosen symmetry-adapted
  to the complete magnetic group $M_{110}$, cf. note (xii) of Table A2
  in~\cite{enio}.
\end{itemize}
\end{enumerate}
\end{table}

\begin{table}[!]
\caption{
Compatibility relations between the Brillouin zone for the fcc space group
$Fm3m$ (225) of paramagnetic CoO and the Brillouin zone for the space
group $I4_1\!/\!cd$ (110) of the antiferromagnetic structure in distorted CoO.
\label{tab:falten225_110}
}
\begin{flushleft}
\centering
\begin{tabular}[t]{cccccccccc}
\multicolumn{10}{c}{$\Gamma (000)$}\\
\hline
$\Gamma^+_1$ & $\Gamma^+_2$ & $\Gamma^-_2$ & $\Gamma^-_1$ & $\Gamma^+_3$ & $\Gamma^-_3$ & $\Gamma^+_4$ & $\Gamma^+_5$ & $\Gamma^-_4$ & $\Gamma^-_5$\\
$\Gamma_1$ & $\Gamma_4$ & $\Gamma_3$ & $\Gamma_2$ & $\Gamma_1$ + $\Gamma_4$ & $\Gamma_2$ + $\Gamma_3$ & $\Gamma_2$ + $\Gamma_5$ & $\Gamma_3$ + $\Gamma_5$ & $\Gamma_1$ + $\Gamma_5$ & $\Gamma_4$ + $\Gamma_5$\\
\hline\\
\end{tabular}\hspace{1cm}
\begin{tabular}[t]{cccccccccc}
\multicolumn{10}{c}{$X (\frac{1}{2}0\frac{1}{2})$}\\
\hline
$X^+_1$ & $X^+_2$ & $X^+_3$ & $X^+_4$ & $X^+_5$ & $X^-_1$ & $X^-_2$ & $X^-_3$ & $X^-_4$ & $X^-_5$\\
$\Gamma_3$ + $\Gamma_5$ & $\Gamma_1$ + 2$\Gamma_4$ & $\Gamma_2$ + $\Gamma_5$ & 2$\Gamma_1$ + $\Gamma_4$ & $\Gamma_2$ + $\Gamma_3$ + 2$\Gamma_5$ & $\Gamma_4$ + $\Gamma_5$ & $\Gamma_2$ + 2$\Gamma_3$ & $\Gamma_1$ + $\Gamma_5$ & 2$\Gamma_2$ + $\Gamma_3$ & $\Gamma_1$ + $\Gamma_4$ + 2$\Gamma_5$\\
\hline\\
\end{tabular}\hspace{1cm}
\begin{tabular}[t]{cccccc}
\multicolumn{6}{c}{$L (\frac{1}{2}\frac{1}{2}\frac{1}{2})$}\\
\hline
$L^+_1$ & $L^+_2$ & $L^-_1$ & $L^-_2$ & $L^+_3$ & $L^-_3$\\
$\Gamma_2$ + $\Gamma_4$ + $\Gamma_5$ & $\Gamma_1$ + $\Gamma_3$ + $\Gamma_5$ & $\Gamma_1$ + $\Gamma_3$ + $\Gamma_5$ & $\Gamma_2$ + $\Gamma_4$ + $\Gamma_5$ & $\Gamma_1$ + $\Gamma_2$ + $\Gamma_3$ + $\Gamma_4$ + 2$\Gamma_5$ & $\Gamma_1$ + $\Gamma_2$ + $\Gamma_3$ + $\Gamma_4$ + 2$\Gamma_5$\\
\hline\\
\end{tabular}\hspace{1cm}
\begin{tabular}[t]{cccc}
\multicolumn{4}{c}{$\Sigma_M (\frac{1}{4}\frac{1}{4}\frac{1}{2})$}\\
\hline
$\Sigma_1$ & $\Sigma_2$ & $\Sigma_3$ & $\Sigma_4$\\
$Z_1$ + $Z_2$ + $Z_5$ & $Z_3$ + $Z_4$ + $Z_5$ & $Z_3$ + $Z_4$ + $Z_5$ & $Z_1$ + $Z_2$ + $Z_5$\\
\hline\\
\end{tabular}\hspace{1cm}
\begin{tabular}[t]{ccccc}
\multicolumn{5}{c}{$W'' (\frac{1}{4}\overline{\frac{1}{4}}\overline{\frac{1}{2}})$}\\
\hline
$W_1$ & $W_2$ & $W_3$ & $W_4$ & $W_5$\\
$Z_5$ & $Z_5$ & $Z_5$ & $Z_5$ & $Z_1$ + $Z_2$ + $Z_3$ + $Z_4$\\
\hline\\
\end{tabular}\hspace{1cm}
\begin{tabular}[t]{ccccc}
\multicolumn{5}{c}{$\Delta_M' (\overline{\frac{1}{4}}\overline{\frac{1}{4}}0)$}\\
\hline
$\Delta_1$ & $\Delta_2$ & $\Delta_3$ & $\Delta_4$ & $\Delta_5$\\
$Z_3$ + $Z_4$ & $Z_1$ + $Z_2$ & $Z_1$ + $Z_2$ & $Z_3$ + $Z_4$ & 2$Z_5$\\
\hline\\
\end{tabular}\hspace{1cm}
\begin{tabular}[t]{ccccc}
\multicolumn{5}{c}{$\Delta_M (\frac{1}{4}0\frac{1}{4})$}\\
\hline
$\Delta_1$ & $\Delta_2$ & $\Delta_3$ & $\Delta_4$ & $\Delta_5$\\
$X_1$ & $X_1$ & $X_1$ & $X_1$ & 2$X_1$\\
\hline\\
\end{tabular}\hspace{1cm}
\begin{tabular}[t]{ccc}
\multicolumn{3}{c}{$\Lambda_M (\frac{1}{4}\frac{1}{4}\frac{1}{4})$}\\
\hline
$\Lambda_1$ & $\Lambda_2$ & $\Lambda_3$\\
$N_1$ + $N_2$ & $N_1$ + $N_2$ & 2$N_1$ + 2$N_2$\\
\hline\\
\end{tabular}\hspace{1cm}
\begin{tabular}[t]{ccccc}
\multicolumn{5}{c}{$W' (\overline{\frac{3}{4}}\overline{\frac{1}{2}}\overline{\frac{1}{4}})$}\\
\hline
$W_1$ & $W_2$ & $W_3$ & $W_4$ & $W_5$\\
$X_1$ & $X_1$ & $X_1$ & $X_1$ & 2$X_1$\\
\hline\\
\end{tabular}\hspace{1cm}
\end{flushleft}
\ \\
\begin{flushleft}
Notes to Table~\ref{tab:falten225_110}:
\end{flushleft}
\begin{enumerate}
\item The upper rows list the representations of the little groups of the
 points of symmetry in the Brillouin zone for $Fm3m$, and the lower rows list
 representations of the little groups of the related points of symmetry in
 the Brillouin zone for $I4_1\!/\!cd$.
 
 The representations in the same column are compatible in the
 following sense: Bloch functions that are basis functions of a
 representation $\bm{D}_i$ in the upper row can be unitarily transformed into
 the basis functions of the representation given below $\bm{D}_i$.
\item The notations of the points of symmetry follow Figures 3.14 and
 3.10 (b), respectively, of \cite{bc}.
\item The notations of the representations are defined in
 Table A4 of~\cite{enio} and Table~\ref{tab:rep_110}, respectively.
\item The compatibility relations are determined as described in great
  detail in \cite{eabf}.
\end{enumerate}
\end{table}

\begin{table}[t]
\caption{
  Compatibility relations between points and lines in the Brillouin
  zone  for the fcc space group
$Fm3m$ (225) of paramagnetic CoO \big(Table {\bf (a)}\big) and in the Brillouin
zone for the space group $I4_1\!/\!cd$ (110) of the antiferromagnetic
structure of distorted CoO \big(Table {\bf (b)}\big), as far as the are useful
for an understanding of Figures~\ref{fig:bandstr225} and~\ref{fig:bandstr110}.   
\label{tab:comp_rel_225_110}
}
\begin{flushleft}
\centering
\begin{tabular}[t]{cccccccccc}
\multicolumn{10}{c}{$\Gamma (000)$}\\
\hline
$\Gamma^+_1$ & $\Gamma^+_2$ & $\Gamma^-_2$ & $\Gamma^-_1$ & $\Gamma^+_3$ & $\Gamma^-_3$ & $\Gamma^+_4$ & $\Gamma^+_5$ & $\Gamma^-_4$ & $\Gamma^-_5$\\
$\Delta_1$ & $\Delta_3$ & $\Delta_4$ & $\Delta_2$ & $\Delta_1$ + $\Delta_3$ & $\Delta_2$ + $\Delta_4$ & $\Delta_2$ + $\Delta_5$ & $\Delta_4$ + $\Delta_5$ & $\Delta_1$ + $\Delta_5$ & $\Delta_3$ + $\Delta_5$\\
\hline\\
\end{tabular}\hspace{1cm}
\begin{tabular}[t]{cccccccccc}
\multicolumn{10}{c}{$\Gamma (000)$}\\
\hline
$\Gamma^+_1$ & $\Gamma^+_2$ & $\Gamma^-_2$ & $\Gamma^-_1$ & $\Gamma^+_3$ & $\Gamma^-_3$ & $\Gamma^+_4$ & $\Gamma^+_5$ & $\Gamma^-_4$ & $\Gamma^-_5$\\
$\Sigma_1$ & $\Sigma_3$ & $\Sigma_4$ & $\Sigma_2$ & $\Sigma_1$ + $\Sigma_3$ & $\Sigma_2$ + $\Sigma_4$ & $\Sigma_2$ + $\Sigma_3$ + $\Sigma_4$ & $\Sigma_1$ + $\Sigma_2$ + $\Sigma_4$ & $\Sigma_1$ + $\Sigma_3$ + $\Sigma_4$ & $\Sigma_1$ + $\Sigma_2$ + $\Sigma_3$\\
\hline\\
\end{tabular}\hspace{1cm}
\begin{tabular}[t]{cccccccccc}
\multicolumn{10}{c}{$X (\frac{1}{2}0\frac{1}{2})$}\\
\hline
$X^+_1$ & $X^+_2$ & $X^+_3$ & $X^+_4$ & $X^+_5$ & $X^-_1$ & $X^-_2$ & $X^-_3$ & $X^-_4$ & $X^-_5$\\
$\Delta_1$ & $\Delta_2$ & $\Delta_3$ & $\Delta_4$ & $\Delta_5$ & $\Delta_2$ & $\Delta_1$ & $\Delta_4$ & $\Delta_3$ & $\Delta_5$\\
\hline\\
\end{tabular}\hspace{1cm}

{\bf (a)}
\ \vspace{.5cm}

\begin{tabular}[t]{ccccc}
\multicolumn{5}{c}{$\Gamma (010)$, $\Gamma (000)$}\\
\hline
$\Gamma_1$ & $\Gamma_2$ & $\Gamma_3$ & $\Gamma_4$ & $\Gamma_5$\\
$F_1$, $\Sigma_1$ & $F_2$, $\Sigma_2$ & $F_1$, $\Sigma_1$ & $F_2$,
$\Sigma_2$ & $F_1$ + $F_2$, $\Sigma_1$ +  $\Sigma_2$\\
\hline\\
\end{tabular}\hspace{1cm}
\begin{tabular}[t]{ccccc}
\multicolumn{5}{c}{$Z (\frac{1}{2}\frac{1}{2}\overline{\frac{1}{2}})$}\\
\hline
$Z_1$ & $Z_2$ & $Z_3$ & $Z_4$ & $Z_5$\\
$F_1$ & $F_1$ & $F_2$ & $F_2$ & $F_1$ + $F_2$\\
\hline\\
\end{tabular}\hspace{1cm}

\begin{tabular}[t]{ccccc}
\multicolumn{5}{c}{$\Gamma (000)$}\\
\hline
$\Gamma_1$ & $\Gamma_2$ & $\Gamma_3$ & $\Gamma_4$ & $\Gamma_5$\\
$\Lambda_1$ & $\Lambda_2$ & $\Lambda_3$ & $\Lambda_4$ & $\Lambda_5$\\
\hline\\
\end{tabular}\hspace{1cm}
\begin{tabular}[t]{ccccc}
\multicolumn{5}{c}{$Z (\frac{1}{2}\frac{1}{2}\overline{\frac{1}{2}})$}\\
\hline
$Z_1$ & $Z_2$ & $Z_3$ & $Z_4$ & $Z_5$\\
$\Lambda_4$ & $\Lambda_2$ & $\Lambda_1$ & $\Lambda_3$ & $\Lambda_5$\\
\hline\\
\end{tabular}\hspace{1cm}

{\bf (b)}

\end{flushleft}
\end{table}

\FloatBarrier

\appendixtitles{no} 

\bibliographystyle{mdpi}
\reftitle{References}

\end{document}